\documentclass[final,sort&compress]{aipproc}
\layoutstyle{6x9}

\usepackage{amsmath,amssymb,amsthm,latexsym}
\usepackage{graphicx}
\usepackage{psfrag}
\usepackage{subfigure}

\newcommand{\av}[1]{{\left<{#1}\right>}}
\newcommand{\var}{\textnormal{var}}
\newcommand{\E}{{\textrm{e}}}
\newcommand{\const}{\textnormal{const.}}

\begin{document}

\title{Perfect Tempering}

\author{M. Daghofer}{
  address={Institute for Theoretical and Computational Physics, TU Graz, Austria}
}

\author{M. Konegger}{
  address={Institute for Theoretical and Computational Physics, TU Graz, Austria}
}
\author{H. G. Evertz}{
  address={Institute for Theoretical and Computational Physics, TU Graz, Austria}
}
\author{W. von der Linden}{
  address={Institute for Theoretical and Computational Physics, TU Graz, Austria}
}

\date{\today}

\begin{abstract}
Multimodal structures in the sampling density (e.g. two competing phases)
can be a serious problem for traditional Markov 
Chain Monte Carlo (MCMC), because correct sampling of the different structures can only be
guaranteed for infinite sampling time. Samples may not decouple from the initial configuration for
a long time and autocorrelation times may be hard to determine.

We analyze a suitable modification \cite{gey:95} of the simulated tempering
idea~\cite{marpar:92}, which has orders of magnitude smaller
autocorrelation times for multimodal sampling densities and which samples all
peaks of multimodal structures according to their weight. The method
generates {\bf exact}, i.\,e. uncorrelated, samples and thus gives 
access to reliable error estimates. {\bf Exact tempering} is applicable to
arbitrary (continuous or discreet) sampling densities and moreover presents a possibility to calculate
integrals over the density (e.g. the partition function for the Boltzmann
distribution), which are not accessible by usual MCMC.
\end{abstract}


\maketitle

\section{Introduction}

Simulated Tempering was introduced in Ref.~\cite{marpar:92},
parallel tempering, also known as Replica Exchange Monte Carlo, in Ref.~\cite{huknem:95,huknem:96} and both have
been widely used (see e.~g. Refs. \cite{kerreh:94,pinwie:98}) to make Markov chain Monte Carlo faster. For an
introduction to both methods see Ref.~\cite{mar:96}.

Propp and Wilson introduced the coupling from the past
(CFTP) method to draw exact samples, i.e. samples which are guaranteed to be
uncorrelated and to obey the desired distribution in Ref.~\cite{propwil:96}. Applications of this
method to continuous degrees of freedom and cluster
algorithms exist, see Refs. \cite{chietal:01} and \cite{hub:98a}.

A small modification of the Simulated
Tempering algorithm likewise allows to obtain uncorrelated samples, see
Ref.~\cite{gey:95}. The first two
sections give an introduction to this procedure and the
resulting error estimates. We then examine the
tempering Markov matrix and the autocorrelation time in
and give indications about the needed
parameters. We investigate the tempering algorithm
for multidimensional continuous distributions and
find a polynomial dependence on the dimension. Finally,
we compare exact sampling with simulated tempering to the CFTP method for the
two dimensional Ising model and find a quadratic dependence on the system
size for simulated tempering, while CFTP needs exponential time for cold but
finite temperatures.

\section{Exact Sampling with Simulated Tempering}\label{sec:exact}

\begin{figure}[htbp]
   \includegraphics[width = 0.8\textwidth]{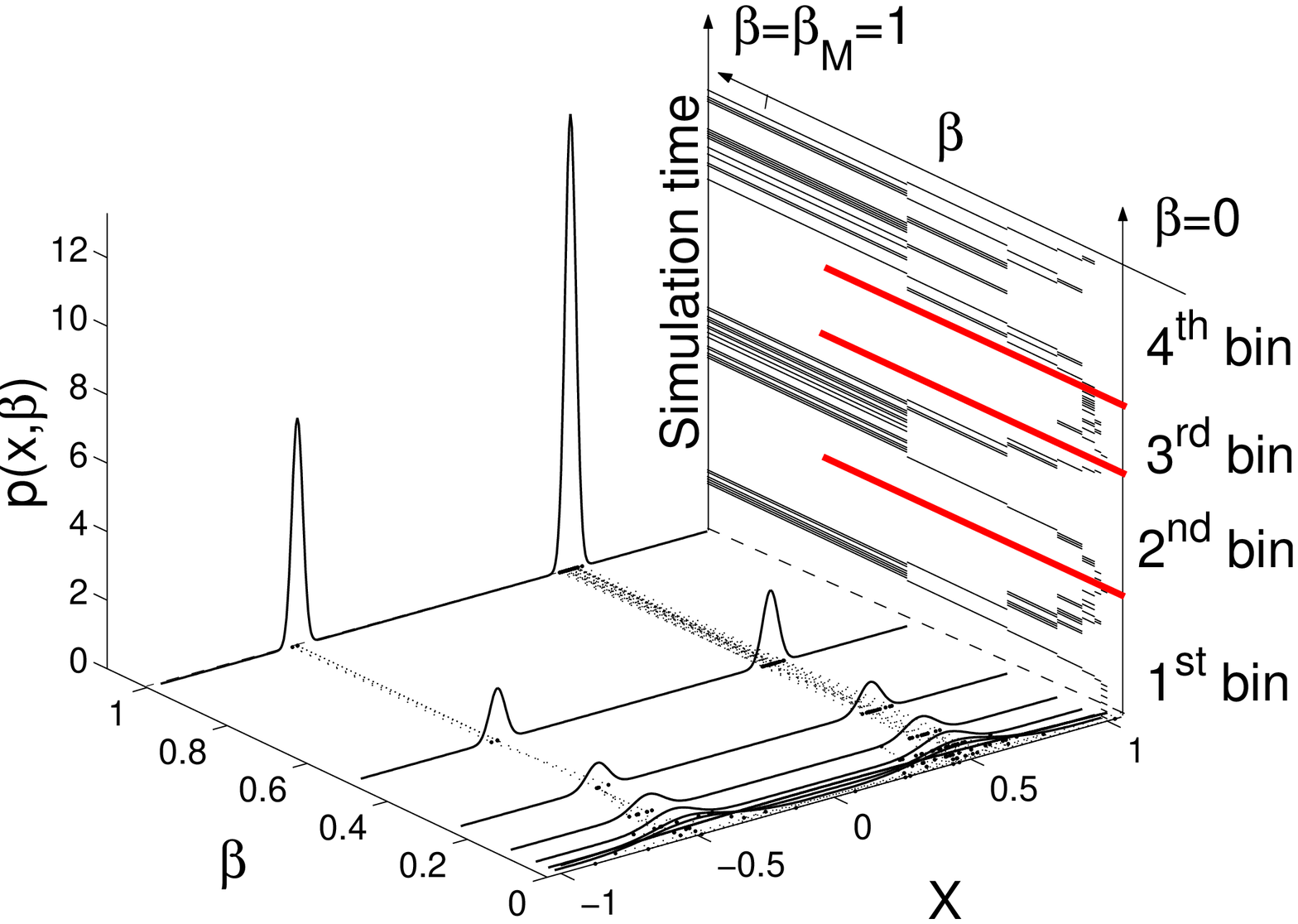}\\
   \caption{Example for a Simulated Tempering run.
     On the `floor', the Markov chain travels through the $\{x,\beta\}$-space,
   the larger dots are the obtained samples, the dotted lines show the way the
   Markov process has taken. Via $\beta_0=0$, the walk reaches both peaks at $\beta_M=1$,
   although no direct tunneling between them occurs.
   The peaks (solid lines) are the probabilities $p(x|\beta)$ for the various discrete
   $\beta$-values. The samples drawn at a certain temperature obey this
   distribution.
   On the right hand `wall', the vertical axis is the time axis of the
   simulation; one sees the wandering of the random walk through the
   temperatures. The thick lines are inserted where the walk reaches
   $\beta_0=0$, i.~e. where an independent exact sample is drawn from
   $p_0=p(x|\beta=0)$ (chosen as a single broad Gaussian peak). At
   these points, the walk forgets its past and a new uncorrelated bin starts.
  \label{fig:walk_bins}}
 \end{figure}

Besides speeding simulations up, Simulated Tempering provides an alternative way to obtain exact samples from arbitrary
probability density functions \cite{gey:95}. Fig.~\ref{fig:walk_bins} shows the principle for a multi-modal distribution $p_1(X)$
consisting of two Gaussians, but it does not depend on the specified
example and can thus be applied to a variety of probability distributions. We want
to draw Exact samples from the distribution $p_1(X)$, which we can not sample directly, where $X$ can be a discrete
or continuous quantity of arbitrary dimension. In order to do so, we introduce an additional parameter $\beta$
and the joint probability $p(X,\beta)=p(X|\beta)p(\beta)$. We have large freedom in choosing
$p(X,\beta)$, for the simulation depicted in fig.~\ref{fig:walk_bins}, we chose:
\begin{equation}\label{eq:jprob}
p(X,\beta_m) = \frac{1}{Z} \frac{1}{Z_m} p_1(X)^{\beta_m}p_0(X)^{1-\beta_m}\;,
\end{equation}
where $Z$ is the overall normalization, $Z_m$ is a constant depending
on $\beta_m$, which determines $p(\beta_m)$. The additional variable 
$\beta$ was allowed to take $M+1$ discrete values $\beta_m$ with $\beta_0 =0$ and $\beta_M=1$.
$p_0(X)$ should be chosen in a way to allow generating Exact samples easily; in our example, it was
a single broad Gaussian peak. Furthermore, its range in $X$-space should be
broad enough to cover all structures of $p_1(X)$. For application to
physical systems, $p_1$ would of course be chosen as $\E^{-E(X)}$ (with
$E(X)$ denoting the energy) and $p_0=\const$, which yields
$p(X|\beta)=\E^{-\beta E(X)}$. In this case, $\beta_M$ is not chosen to be
one, but can take any other value $\beta_{max}$. 

We then do Markov chain Monte Carlo in the $\{X,\beta\}$-space, where we
alternate a couple of sweeps in $X$-space with moves in $\beta$-direction. 
In $\beta$-direction, $\beta_{m'}$ with $m'= m \pm 1$ is
proposed with equal probability and accepted according to the Metropolis scheme.
In $X$-space and for $\beta_m\neq 0$, usual Metropolis updates are
employed. A special case arises for $X$-moves at $\beta_m=0$. In this
case $p(X|\beta=0)\propto p_0(X)$ and we are able to draw a new
exact sample $X'$ distributed according to $p_0(X)$, which gives us
a sample $X'$ \emph{uncorrelated} from $X$.   

An example of the resulting random walk is depicted on the 'floor' of
Fig.~\ref{fig:walk_bins}. Whenever this random walk 
reaches $\beta=0$, a new exact sample from $p_0$ is drawn independent from the current state of the
Markov chain so that the walk forgets its past. The MC time needed for one
exact sample is thus given by the time needed by the Markov chain to travel from $\beta_0=0$ to $\beta_M=1$ and back
again.

A plain MCMC run would instead be trapped in one of the two peaks and
rarely tunnel to the other. Repeating several plain MCMC runs and taking
their average would give the wrong expectation value $\bar x=0$, because
the different weight of the peaks would not be accounted for.

\section{Expectation values and error estimates}
\label{sec:temp_exp_err}

As the $\{X,\beta\}$-samples obtained by the simulation obey $p(X,\beta)$, the $X$ drawn at a given
temperature $\beta_m$ obeys $p(X|\beta_m)$. Expectation values for $\beta_M=1$ are therefore calculated from all
(correlated and uncorrelated) samples obtained at a given temperature:
\begin{equation}\begin{split}\label{eq:mean}
  \bar X(\beta_M=1) &= \frac{1}{N_M} \sum\limits_{j=1}^{N_M} X_{j} = \frac{1}{N_M}
  \sum\limits_{i=1}^{N_{M,ind}} \sum\limits_{\nu=1}^{N_i} X_{i,\nu}\\
  \langle \bar X(\beta_M=1) \rangle &= \frac{1}{N_M} \sum\limits_{j} \langle
  X_{j} \rangle = \langle X \rangle
\end{split}\end{equation}
The $X_j$ are the measurements obtained at the desired temperature $\beta_M=1$, their index $j$ was
broken into $i$ and $\nu$ with $i$ denoting the independent and uncorrelated bins and $\nu$
labeling the correlated measurements within one bin, see Fig.~\ref{fig:walk_bins}. $N_{M,ind}$ is
the number of independent bins which contain at least one sample drawn at $\beta_M$, and $N_i$ the
number of measurements within the $i$-th bin. $N_M=\sum_{i=1}^{N_{M,ind}} N_i$ is the total number
of times the simulation has visited the desired temperature $\beta_M=1$. A bar
denotes the sample mean obtained in the Monte Carlo run, $\av{\dots}$ denotes an
expectation value over all samples. The sample mean is obviously unbiased.

It is worth noting that measuring the bin averages does not give the same result, because the
probability for a move in $\beta$-direction, and thus the number of measurements ($N_i$) taken in a bin
before the walk returns to $\beta=0$, is a random variable and depends on the current sample $X$:
\begin{equation}\label{eq:bin_mean}
    \langle \frac{1}{N_{ind}}\sum\limits_{i=1}^{N_{ind}}
    \frac{\sum_{\nu=1}^{N_i}X_{i,\nu}} {N_i} \rangle \neq
    \langle \frac{1}{N_{ind}} \sum_{i=1}^{N_{ind}}
    \frac{\sum_{\nu=1}^{N_i} X_{i,\nu}} {\bar N_{bin}}
    \rangle 
    = \langle \frac {\sum_{i,\nu} X_{i,\nu}}
    {\underbrace{N_{ind} \bar N_{bin}} _{=N}} \rangle = \langle X \rangle
\end{equation}
Here,  $\bar N_{bin} = \frac{1}{N_{ind}} \sum_{i=1}^{N_{ind}} N_i$ is the average number of
measurements per bin. For the same reason, taking only the first sample of each bin does not give
correct results. For a multi-modal $p_1(X)$ with a different height (and/or width) of the peaks as in
Fig.~\ref{fig:walk_bins}, the Markov Chain may visit the smaller peak very often, but it will stay
at the larger one longer.

The independent samples provide a way to analyze correlations and to calculate reliable error
estimates \cite{gey:95}. When calculating the variance of the estimate $\bar X$, the new labels $i$ and $\nu$
become useful as it is now important to distinguish between correlated and uncorrelated samples:
\begin{equation}\begin{split}
\label{eq:squ_mean}
    \langle \bar{X}^2\rangle & =
    \frac{1}{N^2} \sum\limits_{i,j} \langle \sum\limits_{\nu=1}^{N_i}
    \sum\limits_{\mu=1}^{N_j} X_{i,\nu} X_{j,\mu} \rangle = \\
    & = \frac{1}{N^2}
    \sum\limits_{i,j} \langle \sum\limits_{\nu=1}^{N_i}
    \sum\limits_{\mu=1}^{N_j} \Delta X_{i,\nu} \Delta X_{j,\mu} \rangle + 
    \frac{2\langle X\rangle}{N^2}
    \langle \underbrace{\sum\limits_j N_j}_{=N} \sum\limits_{i,\nu}
     \Delta X_{i,\nu} \rangle +
     \frac{\langle X\rangle^2}{N^2}
    \langle \underbrace{\sum\limits_{i,j} N_i N_j}_{=N^2} \rangle=\\
    & = \frac{1}{N^2}
    \sum\limits_{i} \langle \sum\limits_{\nu,\mu=1}^{N_i}
    \Delta X_{i,\nu} \Delta X_{i,\mu} \rangle +
    \frac{1}{N^2}\sum\limits_{i\neq j} \underbrace
    {\langle \sum\limits_{\nu=1}^{N_i}\Delta X_{i,\nu}\rangle }
    _{=0}\underbrace
    {\langle \sum\limits_{\mu=1}^{N_j}\Delta X_{j,\mu}\rangle }
    _{=0} +
     \frac{2\langle X\rangle}{N}
    \underbrace{ \langle \sum\limits_{i,\nu} \Delta X_{i,\nu} \rangle}
    _{=0} + \langle X\rangle^2
\end{split}\end{equation}

where $\langle \sum_{\nu,\mu}\Delta X_{i,\nu} \Delta X_{j,\mu}\rangle  = \langle \sum_{\nu} \Delta
X_{i,\nu}\rangle \langle \sum_{\mu} \Delta X_{j,\mu}\rangle$ for $i\neq j$, because the
measurements are from different bins, $\langle \sum_{\nu,\mu=1}^{N_i} \Delta X_{i,\nu} \Delta
X_{i,\mu} \rangle$ is independent of $i$, because all bins are
equivalent. From Eq.~\ref{eq:squ_mean}, it follows for the variance
\begin{equation}\label{eq:err_est}
        \var(\bar X)=\langle\bar{X}^2\rangle -\langle\bar X\rangle^2
        =\langle \bar{X}^2\rangle -\langle X \rangle ^2
        =\frac{N_{ind}}{N^2} \langle \sum\limits_{\nu,\mu=1}^{N_i}
         \Delta X_{i,\nu} \Delta X_{i,\mu} \rangle\;.
\end{equation}
The unknown expectation value $\langle \sum_{\nu,\mu=1}^{N_i} \Delta X_{i,\nu} \Delta X_{i,\mu}
\rangle$ is estimated from the Monte Carlo run, thus $\langle \sum_{\nu,\mu=1}^{N_i} \Delta
X_{i,\nu} \Delta X_{i,\mu} \rangle_{\textnormal{est}} \approx \frac{1}{N_{ind}}
\sum_{i=1}^{N_{ind}} \sum_{\nu,\mu=1}^{N_i} \Delta X_{i,\nu} \Delta X_{i,\mu}$.
However, the variance depends on the determination of the above
expectation value, so it can only be correct, if all modes of $p_1$
have been sampled sufficiently. Similar formulae can be derived for the
expectation values and error estimates of more complex observables (e.g. of
the covariance), where correlations between the measured parameters can thus
be taken into account. 

\section{Behavior in one dimension}
\label{sec:temp_Markov_Matr}

Although nobody would think of using Monte Carlo simulation for one
dimensional problems, as much more efficient approaches are available,
it is interesting to examine the Markov matrix for a Simulated
Tempering simulation in the two-dimensional
$X$-$\beta$-space with discretized $X$.
The probability density $p_1(x)$ for $\beta=1$ was chosen to consist of two
Gaussians well separated from each other and $p_0(x)$ was chosen to be constant.
For Simulated Tempering, the number of $\beta$-slices was varied from
two (just $\beta=0$ with $p(X|\beta=0)=p_0$ and $\beta=1$ with
$p(X|\beta=1)=p_1$) to five. The intermediate $\beta$-values were
chosen so as to give approximately the same transition rate between all pairs of
adjacent $\beta$-values. Autocorrelation and thermalization are largely determined by the
second largest eigenvalue ($e_2$) of the Markov matrix, i.\,e. the one with magnitude closest to one.
The autocorrelation time was approximately calculated as $\tau_{AC}\approx1/(1-|e_2|)$.

\begin{figure}
\includegraphics[width=0.6\textwidth]{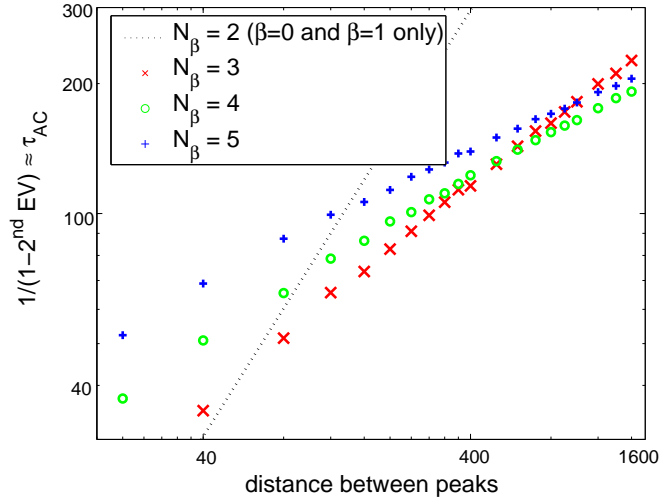}
\caption{Autocorrelation
  time for the Simulated Tempering algorithm in 1D 
depending on the distance between the two peaks for two to five
$\beta$-slices. The distance is measured in multiples of the width
$\sigma$ of the Gaussian.\label{fig:mini_sim}}
\end{figure}
 
Fig.~\ref{fig:mini_sim} shows this autocorrelation time as a function
of the distance of the two peaks. One sees that more $\beta$-slices
become necessary as the distance increases. For plain Markov chain
Monte Carlo in the one-dimensional discrete $X$-space, the
autocorrelation time far exceeded the range plotted in
Fig.~\ref{fig:mini_sim} even for a distance of $d=12$ ($\tau_{AC}
\approx 2.6499e+03$) and its calculation is numerically instable for
larger distances.

The columns of the Tempering Markov matrix which correspond to
$\beta=0$ are identical, which means just that whenever the current
state of the chain is at $\beta=0$, the outcome of the next move will
not depend on the current position in $X$-space.

\section{Parallel Tempering}
\label{sec:partemp}

Another method similar to Simulated Tempering Parallel Tempering, also called
Exchange Monte Carlo, see Refs.~\cite{mar:96, huknem:95}. In this method, we have
$M$ copies of $X$ at the $M$ values for $\beta$.
Instead of the
space $\{X,\beta_m\}$ as in Simulated Tempering, we now consider the
product space $\{X_0,X_1,\dots,X_m,\dots,X_M\}$ where the
configuration $X_m$ is at the temperature $\beta_m$. At every $\beta_m$, there is
\emph{exactly one} configuration $X$, denoted by $X_m$ and obeying the
distribution $p(X|\beta_m)$.
The probability for the total product space is given by the product of
the individual probabilities.
We now do Markov chain Monte Carlo again with this product
probability.
In $X$-space, Metropolis Monte Carlo updates are performed for all $\beta$s
independently. New configurations $X^{'}_{m}$ are obtained with the usual
Metropolis random walk for $\beta>0$, while a new sample is drawn directly from $p_0(X)$
for $\beta=0$. 
Alternated with the updates in $X$-space, Metropolis moves to swap configurations $X_m$ and $X_{m+1}$ at adjacent
$\beta$-values are performed.

During the Monte
Carlo run, all $X-m$ will eventually get swapped to $\beta=0$, where a new
sample is drawn. This time, however, the random walk does not
completely forget its past, which can be inferred from the Markov
matrix for a similar toy situation as for Simulated Tempering
above. Suppose, we have three $\beta$-values $\beta_0=0,\;\beta_1,\;
\beta_2=1$, and the following temperature swaps occur in the Markov
chain Monte Carlo:

\begin{equation*}
  \begin{matrix} 2 \\ 1 \\ 0 \end{matrix}\ \rightarrow\
  \begin{matrix} 2 \\ \tilde 0 \\ 1 \end{matrix}\ \rightarrow\
  \begin{matrix} \tilde 0 \\ 2 \\ \tilde 1 \end{matrix}\ \rightarrow\
  \begin{matrix} \tilde 0 \\ \tilde 1 \\ 2 \end{matrix}\ \rightarrow\
  \begin{matrix} \tilde 0 \\ \tilde 1 \\ \tilde 2 \end{matrix}\;,
\end{equation*}

where a tilde means that an exact sample is drawn from $p_0$. All
Configurations have now been at $\beta=0$, but the columns of the
matrix corresponding to the above sequence of swaps are still not
equal, which means that the current state of the Markov chain still
depends on its initial state. However, these correlations are small
after an initial thermalization and autocorrelation times are short.

\section{Needed Parameters}
\label{sec:adjust}

In order to do Simulated or Parallel Tempering, we have to adjust the values
for the $\beta_m$ and the $Z_m$, see Eq.\eqref{eq:jprob}.
The $\beta_m$-values have to be dense enough to give a
considerable overlap of $p(X|\beta_m)$ and $p(X|\beta_{m\pm 1})$.
On the other hand, we want to have as few $\beta$-values as
possible between $\beta=1$ and $\beta=0$. The $\beta$-values can be
adjusted in a Parallel Tempering prerun, where a new value is inserted
whenever the swapping rate between adjacent $\beta$s is too low. 

The ideal $Z_m$ needed for Simulated Tempering would make all $\beta$-values equally
likely. This prevents the Markov chain from spending too much time
at on single temperature and thus speeds travel from $\beta=0$ to
$\beta=1$ and back again. This leads to:
\begin{equation}
  Z_m \propto \int\limits_{X} dX  p_1(X)^{\beta_m} p_0(X)^{1-\beta_m}\nonumber
\end{equation}
For physical systems, the weight $Z(\beta)$ gives the partition function,
which can only be determined in terms of 
$Z(\beta=0)$. For problems in data analysis, it is the model evidence,
i.e. the probability for the chosen model integrated over all possible
parameter values. The weights can be obtained from the visiting frequency for
the $\beta$-values in Simulated Tempering preruns, but this is rather
difficult, because they may differ by orders of magnitude. They
are not needed for Parallel Tempering, where they cancel out, but the
integral can still be calculated with a procedure similar to
thermodynamical integration, see Ref.~\cite{pinwie:98}. With the random
samples produced at $\beta_m$, we can estimate $Z_{m+1}$ for $\beta_{m+1}$:
\begin{equation}
  \frac{Z_{m+1}}{Z_m}  = \langle \frac {p_1(X)^{\beta_{m+1}} p_0(X)^{1-\beta_{m+1}}} 
  {p_1(X)^{\beta_m} p_0(X)^{1-\beta_m}} \rangle_{\beta_m}\;,
\end{equation}
where $\langle \dots\rangle_{\beta_m}$ denotes an expectation value
calculated at $\beta_m$.
The integral $Z(\beta=1)$ is the product of all the measured ratios:
\begin{equation}
Z(\beta=1) = Z_M = Z_0 \cdot \prod \limits_{m=0}^{M-1}
\frac{Z_{m+1}}{Z_m}\;.
\end{equation}
Care must be taken in evaluating this quantity, because the
configurations are interchanged between $\beta$-values and the 
measurements obtained for the different $\beta$-values are therefore
heavily correlated and the same applies to using parallel tempering data for
multihistogrmming (Ref.~\cite{ferswe:89}), as was proposed in Ref.~\cite{mitoka:00}.

\section{Behavior in higher dimensions}
\label{sec:temp_D}

\begin{figure}
  \centering
  \includegraphics[width=0.45\textwidth]{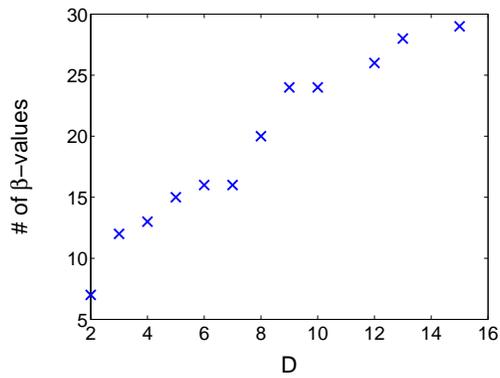}
  \caption{Number $N_\beta$ of needed $\beta$-values needed for various
    dimensions D.\label{fig:d_beta}}
\end{figure}

In this section, we examine the behavior of the Tempering
algorithm in higher dimensions. We chose $p_0$
as one single broad Gaussian with width $\sigma_0=1$ centered at $X=0$ and the wanted probability $p_1$
consisted of two Gaussians of width $\sigma=0.04$ centered at
$X=(0.3, 0.3, \dots)$ and $X=(0.8, 0.8, \dots)$,
which were multiplied by 5000, so as to
yield a norm $n=10\,000$. 100 sweeps were performed between
$\beta$-moves, the $\beta_m$ and $Z_m$ were adjusted in a parallel tempering
prerun. The geometric mean was used to insert new slices, except for finding
the second lowest $\beta_1>0$, where the old value was just divided by a
constant, if the swapping rate was too low.
As the number of needed $\beta$ values $N_\beta$ depends on the logarithm of
the ratio of the volumes of $p_1$ and $p_0$
$N_\beta\propto-\log\left((\sigma_1/\sigma_0)^D\right)$, the dependence on the dimension
$D$ is expected to be  
linear, which is indeed approximately the case, as can be seen in Fig.~\ref{fig:d_beta}.

\begin{figure}
  \centering
  \includegraphics[width=0.45\textwidth]{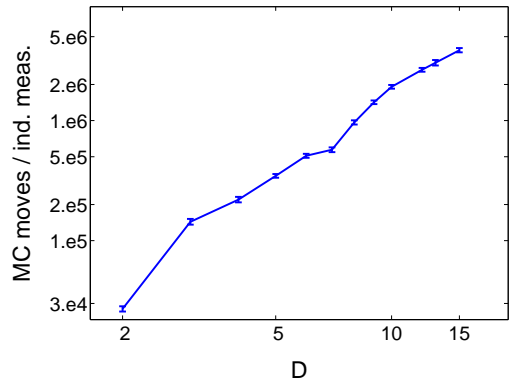}
\caption{Number of needed MC
  updates per independent sample for various dimensions $D$.\label{fig:d_moves}}
\end{figure}

\begin{figure}
  \centering
  \includegraphics[width=0.45\textwidth]{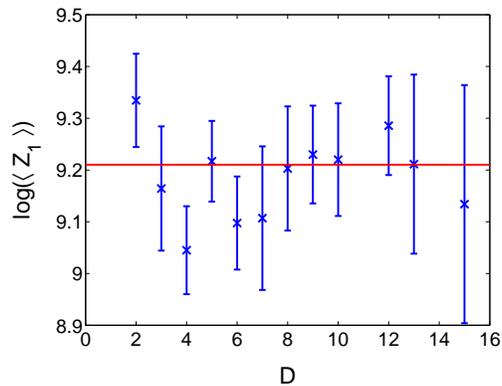}
\caption{Logarithm of the obtained norm $\log(Z(\beta=1))$ for 
  various dimensions $D$.\label{fig:d_av_Z}}
\end{figure}

Figure \ref{fig:d_moves} shows
the number of MC updates needed for one independent sample. One sees
that the increase in needed samples with the dimension of the problem
approximately obeys a power law in contrast to the behavior found for the
CFTP algorithm (Ref.~\cite{chietal:01}), which has an exponential dependence on the
dimension and generally similar performance as the rejection method, see Ref.~\cite{konegger_dipl}. For all
presented dimensions, the results for the norm were consistent with
the errorbars (see Fig. \ref{fig:d_av_Z}) and likewise the average for $X$, i.~e. the simulation
found both peaks. 

\section{Application to the Ising Model}
\label{sec:temp_ising}

\begin{figure}[htbp]
  \centering
  \includegraphics[width = 0.45\textwidth]{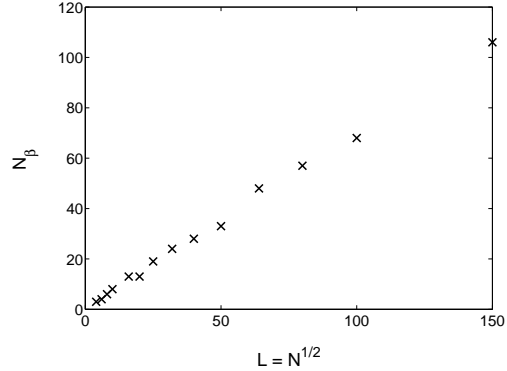}\label{fig:NL_N}
  \caption{Number $N_\beta$ of needed $\beta$-values needed for the two
  dimensional Ising model depending on the linear
  system size $L=\sqrt{N_{spins}}$.}
\end{figure}

Although other choices could be more promising \cite{katzgr:06}, we chose
constant transition (Simulated Tempering) 
and swapping rates (Parallel Tempering, see Ref.~\cite{huknem:95}) between
adjacent temperatures. This leads to: 
\begin{equation}
  \label{eq:delta_beta}
  \frac{1}{(\Delta\beta)^2} \propto \frac{d}{d\beta}\langle E\rangle_\beta
  \propto \langle E^2 \rangle_\beta - \langle E \rangle_\beta^2 =
  \frac{C_v}{k_B\beta^2}\;,
\end{equation}
where $E$ is the energy, $C_v$ the specific heat and $k_B$ the Boltzmann constant.
This relation shows that we need denser $\beta$-values where $C_v$ is large, i.~e. near a (second
order) phase transition. As the specific heat is small for low temperatures again, further cooling
below the phase transition is easy. The specific heat is an extensive quantity, the number of
needed temperatures is therefore expected to grow as $N_\beta\propto \sqrt{N_{spins}}$ with $N_{spins}$ the number
of spins. In this case, we used the arithmetic men to insert new
$\beta$-values. As expected, the number of needed temperatures scales proportional to the
linear system size $N_\beta \propto L=\sqrt{N_{spins}}$, see fig.~\ref{fig:NL_N}.

\begin{figure}[htbp]
  \centering
  \includegraphics[width = 0.45\textwidth]{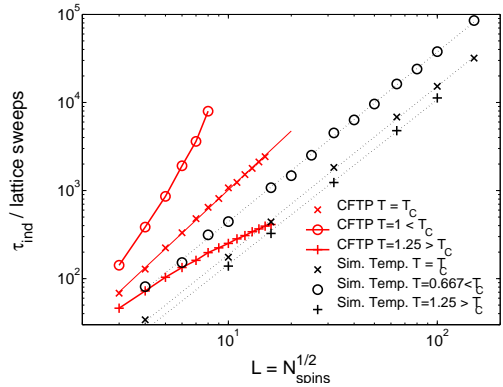}\label{fig:sim_PW}
  \caption{Time per independent sample for the two dimensional Ising model depending on the linear
    system size $L=\sqrt{N_{spins}}$. Solid lines:  CFTP Method and single spin flip algorithm
    Dotted lines: Simulated Tempering and Swendsen Wang algorithm}
\end{figure}

Figure \ref{fig:sim_PW} shows the time per independent sample for Exact
Tempering and for the Coupling From The Past (CFTP) method with single spin
flips introduced by Propp and Wilson in Ref.~\cite{propwil:96}. For CFTP, the
time needed for an independent sample grows on a logarithmic scale with the
system sizes for temperatures above the critical $T_C$, obeys a power law at
$T_C$ and grows exponentially for the ordered phase below $T_C$. There are
CFTP schemes for the Swendsen Wang algorithm, but its runtime scales
exponentially with the system size except at $\beta=0, T=\infty$ and it is
\emph{much} slower than the single spin flip algorithm around $T_C$, see Ref.~\cite{konegger_dipl}.

Exact Tempering with the Swendsen Wang algorithm, on the other hand, gives
linear scaling for all temperatures. Although this is slower than CFTP for high
temperatures, one does in fact get the high temperature results
for free, because they are sampled anyway in a Tempering run for low
temperatures. The critical exponent for Exact Tempering is two for both the
Swendsen Wang ($1.92 \pm 0.06$ at the lowest temperature) and the Wolff
($1.98 \pm 0.07$) algorithm. The reason for this is, that the 
time for an independent sample is determined by the number of steps needed to
go from $\beta=0$ to $\beta=\beta_{max}$ and back again and not by the
algorithm used for the spin updates. This random walk in the temperatures
scales proportional to the square of their number $N_\beta$, which gives

\begin{equation}
\tau_{ind}\propto N_\beta^2\propto L^2 = N_{spins}\;.
\end{equation}

This dependence $\tau_{ind}\propto N_{spins}$ breaks down for the single spin
flip algorithm. Exact Tempering then becomes much slower, because the spin
configurations at the critical temperature cannot be sampled as
efficiently. This effect becomes more severe for first order transitions,
where the algorithm does not manage the transition from the disordered to the
ordered phase at $T_C$. `Tempering' of a model parameter which carries the
transition from first to second order might then be a solution. This was
introduced in Ref. \cite{kerweb:93_1} for the Swendsen-Wang algorithm
applied to the Potts model, where the variable `tempering' parameter was the
number of states $q$. Exact Tempering is also applicable to this variant,
because the percolation problem for $q=1$ can be sampled exactly.

\section{Conclusions}
\label{sec:conclusions}
This paper provides a discussion of Exact Sampling with Simulated Tempering
\cite{gey:95} and compares it to Exact Sampling via the Propp-Wilson
method \cite{propwil:96}. The former is found to be advantageous in most cases.
Simulated Tempering provides a way to draw exact, i.e. completely
uncorrelated samples from arbitrary distributions in high
dimensions. The peaks of multimodal densities are sampled with their
respective weights. The parameters $\beta_m$ and $Z_m$ needed for the Simulated Tempering run can
be adjusted in a Parallel Tempering prerun. While the Parallel Tempering
algorithm itself does not provide perfectly uncorrelated samples, its
autocorrelation time is small. For practical purposes, it is a robust
alternative, because it does not need the parameters $Z_m$. Both methods allow to
calculate the integral over the probability density, i.e. partition functions
or model evidences.
\section{Acknowledgment}

This work has been supported by the Austrian Science Fund (FWF), project
no.\ P15834-PHY. This document has been typeset using the \AmS-\LaTeX\ packages.


\end{document}